# A Monte Carlo study of the fluctuations in Xe electroluminescence yield: pure Xe vs Xe doped with $CH_4$ or $CF_4$ and planar vs cylindrical geometries

J. Escada[a], T.H.V.T. Dias[a,*], F.P. Santos[a], P.J.B.M. Rachinhas[a,b], C.A.N. Conde[a] and A.D. Stauffer[c]

[a] *Laboratório de Instrumentação e Física Experimental de Partículas,
Rua Larga, Coimbra 3004-516, Portugal*

[b] *Serviço de Radioterapia dos Hospitais da Universidade de Coimbra,
Coimbra 3000-075, Portugal*

[c] *Physics and Astronomy Department, York University,
Toronto M3J1P3, Canada*
*E-mail*: fteresa@fis.uc.pt

ABSTRACT: Monte Carlo simulation is used to investigate the fluctuations in the Xe proportional electroluminescence (EL) yield $\mathcal{H}$ (also known as secondary scintillation) produced by sub-ionization primary electrons drifting in the gas under appropriate electric fields, comparing pure Xe gas with Xe doped with $CH_4$ or $CF_4$. The work is modeled on gas detectors of the gas proportional-scintillation type, where amplification is achieved through the production of EL under a charge-multiplication free regime. The addition of the molecular gases to Xe reduces electron diffusion, a desirable effect in large size detectors where primary electrons drift across a long absorption/drift region. However, the presence of the molecules reduces $\mathcal{H}$ and increases its fluctuations. In the case of $CF_4$, the effects are very strong due to significant electron attachment in the EL field range, ruling out $CF_4$ as an acceptable additive. The addition of $CH_4$ affects $\mathcal{H}$ and its fluctuations to a much lower extent, and $CH_4$ concentrations lower than ~1% may be an appropriate choice. In addition, Monte Carlo calculations in pure Xe under cylindrical geometry in a regime below charge multiplication have shown that fluctuations in the EL yield $\mathcal{H}$ are an order of magnitude higher than for planar geometry. For both geometries, though, the fluctuations have a negligible effect on the energy resolution, and variations of the anode radius in cylindrical geometry or grid parallelism in planar geometry may be a more significant cause of concern.

---

[*] Corresponding author.

# Contents

## 1. Introduction

Pioneering work by Conde and Policarpo [1]-[4] first established the working principle of the so called Gas Proportional-Scintillation Counters (GPSC). In this type of noble gas filled detectors, a very good energy resolution can be achieved because the amplification process is not based on avalanche multiplication of the primary electrons as in gas proportional ionization counters (GPIC), but on the stepwise production of electroluminescence (EL) photons by the primary electrons (secondary scintillation, essentially in the VUV). The electrons drift in a confined region under accelerating electric fields just high enough to excite but not ionize the gas atoms, giving a high EL yield per electron with very low fluctuations. In Xe, this means density reduced electric fields from $E/N$ ~3 to ~17 Td[1], the range that will be examined in the present work, where in a standard planar field geometry the EL yield is known to have an approximate linear increase, as established in [3]. Under these circumstances, Xe filled GPSCs in particular can achieve energy resolutions approaching ultimate intrinsic values and typically half of a GPIC, because the high excitation efficiency for Xe by electron impact leads to high EL yields with very low fluctuations, which are negligible compared with the fluctuations in the number of primary electrons produced per event (measured by the Fano factor [5], [6]). The high excitation efficiency for Xe is related to the very low energy losses in elastic collisions and the absence of vibrational modes which occur if molecules are present. Other relevant references on noble gas GPSCs, noble gas EL emission and the scintillation mechanism are [7]-[22].

    Xenon gas detectors are widely used in applications with a large range of energy requirements, from medical instrumentation to astrophysics and high energy particle physics, due to its high atomic number, high ionization and scintillation yields, and stability over a wide range of temperatures. In particular, experiments which require accurate energy and position resolution are increasingly relying on Xe EL as the amplification method. These include large high-pressure detector experiments searching for 0-ν ββ-decay or rare dark matter events [23]-

---

[1] $N$ is the atomic number-density and 1 Townsend = $10^{-17}$ V cm$^2$

[28]. The proposed NEXT experiment [23]-[26], for example, will be using Xe EL in a Xe gas time projection chamber at ~10 atm.

However, two issues are of major concern when using Xe EL in large detectors, as in the NEXT experiment.

On one hand, electron drift velocities in Xe are low and diffusion coefficients are high, and these are important drawbacks when the primary electron cloud must drift long distances towards the anode. To overcome this problem, a low percentage of some light molecular gases may be added to improve the drift parameters [29]-[31], as discussed in Section 3.1.

On the other hand, in large gas detectors the standard planar geometry (uniform electric field) for the EL region presents technical problems, because accurate grid parallelism is difficult to achieve over large areas, compromising electric field uniformity and thus increasing the EL yield fluctuations. Cylindrical geometry may be an alternative choice, since EL will be produced near properly biased anode wires which can be built with radii of superior accuracy.

These two problems have encouraged the present Monte Carlo EL simulation study along two lines. Firstly, in Section 3.2, we obtain EL yields and their fluctuations in planar geometry when Xe is doped with $CH_4$ or $CF_4$, and compare with the results in pure Xe. Secondly, in Section 3.3 we compare EL fluctuations in planar and cylindrical geometries in pure Xe, considering identical EL outputs at equivalent scintillation field settings in both geometries.

While the addition of $CH_4$ or $CF_4$ to Xe may reduce electron diffusion and increase electron drift velocity due to a cooling of the electrons by the vibrational excitation of the molecules [29]-[34], the EL yield in these mixtures becomes lower and exhibits larger fluctuations than in pure Xe. In this work we search for gas mixtures that represent a compromise between desirable low electron diffusion coefficients and the high-yield, low-fluctuation EL which is characteristic of Xe gas at pressures above a few hundred Torr.

The intrinsic energy resolution $R_{int}$ of a detector where amplification relies on the EL produced in a secondary scintillation region under electric fields below charge multiplication threshold is given (FWHM) by [4]-[6], [12]

$$R_{int}^2 = 2.35^2 \ (1/n) \ (F + J/\mathcal{H}) \tag{1}$$
$$= 2.35^2 \ (1/n) \ (F + Q)$$

where $n$ is the number of primary electrons produced in the gas per absorbed radiation event in the detector absorption/drift region, the Fano factor $F$ is the relative variance $F = \sigma_n^2/n$, $\mathcal{H}$ is the number of EL photons produced per single primary electron drifting across the secondary scintillation region and $J = \sigma_\mathcal{H}^2/\mathcal{H}$ is the corresponding relative variance.

For planar geometry, i.e., under uniform electric fields, $J$ in pure Xe is very low and the term $Q=J/\mathcal{H}$ is known to be negligible [12] compared with the Fano factor ($F_{Xe} \sim 0.17$ [5], [6], [16]), resulting in energy resolutions which may approach the ultimate intrinsic value $R_{int}=2.35 (F/n)^{1/2}$. The presence of molecular additives (or impurities in general), though, will cause a degradation of the energy resolution, essentially because $\mathcal{H}$ will become lower and the fluctuation parameters $J$ and $Q=J/\mathcal{H}$ will become larger as shown later.

For cylindrical geometry, the reduced electric field at a distance $r$ from the anode axis is given by $E/N=K/(Nr)$. The relevant parameters are the reduced anode voltage $K=V_a/\ln(r_c/r_a)$ and the reduced anode radius $Nr_a$ [35], where $V_a$ is the anode voltage, $r_c/r_a$ the cathode-to-anode radii ratio and $N$ the gas number density. The electric field increases rapidly near the anode, but if we want to work in the EL regime, the field $E(r_a)/N$ at the anode surface (as the uniform field in planar geometry) must not rise above the electron multiplication threshold in the gas, in order to avoid the large fluctuations inherent to charge multiplication and the consequent degradation of energy resolution. In the present work we will show that cylindrical geometry brings additional fluctuations to EL in Xe when compared to planar geometry, but this is not expected to be a problem because, as for planar geometry, $Q$ will remain a negligible term in Eq. 1.

## 2. Monte Carlo simulation

The Monte Carlo simulation models used in this work have been described in more detail in earlier publications [13]-[18], [32]-[36].

The cross-sections for the electron scattering by Xe atoms and by $CH_4$ and $CF_4$ molecules used in the present simulations are shown in Fig. 1. For Xe they are described in [18] and for $CH_4$ and $CF_4$ in [32]-[34]. They were tested by comparing the Monte Carlo drift parameters with available measured data in the pure gases and in Xe-$CH_4$ and Xe-$CF_4$ mixtures [32]-[34].

The electron drift parameters, and the EL yields and their fluctuations are calculated by following the drift of the electrons along their successive free paths and multiple elastic, inelastic (and superelastic in the case of $CF_4$) collisions with the atoms and molecules in the gas, according to the corresponding electron scattering cross-sections. In the Xe-$CH_4$ and Xe-$CF_4$ mixtures, the history of an electron may end prematurely if the electron is attached to a molecule before reaching the anode, thus ending its contribution to the EL yield before crossing the whole scintillation gap.

In pure Xe, it is assumed that, for every Xe atom that is excited by electron impact, a $Xe_2^*$ excimer is formed in 3-body collisions with neutral Xe atoms, and a VUV scintillation photon is emitted from the characteristic excimer continuum (centered at 7.2 eV, 172 nm for pressures above a few hundred Torr) [7]-[10].

In Xe-$CH_4$ and Xe-$CF_4$ mixtures, though, quenching of the EL emission must be taken into account, since deactivation channels involving 2- and 3-body collisions of the Xe excited atoms with the molecules compete with the formation of the $Xe_2^*$ excimers, the scintillation emitting species. The direct precursors to the excimers are the lowest resonant $Xe(1s_4)$ and metastable $Xe(1s_5)$ states, and quenching of these two states in collisions involving molecules are investigated in [37]-[44]. The next paragraph summarizes the processes included in the simulation for the deactivation in the mixtures of the Xe states excited by electron impact.

The (few) levels that are excited above $Xe(1s_4)$ and $Xe(1s_5)$ are assumed to decay down to one of these two lower states with equal probability. The $Xe(1s_4)$ and $Xe(1s_5)$ states disappear at rates $k_1=1.47 \cdot 10^{-31}$ $cm^{-6}s^{-1}$ and $k_2=8.10 \cdot 10^{-32}$ $cm^{-6}s^{-1}$, respectively, in 3-body collisions with two Xe atoms to give the $Xe_2^*$ excimers, which then emit a scintillation photon. Quenching occurs when, alternatively to the formation of the excimers, both $Xe(1s_4)$ and $Xe(1s_5)$ are deactivated either in 2-body collisions with $CH_4$ at a rate $k_3=3.15 \cdot 10^{-10}$ $cm^{-3}s^{-1}$ and with $CF_4$ at a rate $k_4=2.40 \cdot 10^{-13}$ $cm^{-3}s^{-1}$, or in 3-body collisions with Xe and $CH_4$ at rate $k_5=2.5 \cdot 10^{-29}$ $cm^{-6}s^{-1}$ and with Xe and $CF_4$ at rate $k_6=1.9 \cdot 10^{-32}$ $cm^{-6}s^{-1}$.

The rate constants are based on various sources as follows. Rates $k_1$ and $k_2$ are taken from [45]. Rates $k_3$ and $k_4$ for $Xe(1s_5)$ are taken from [39] and are assumed to apply also to $Xe(1s_4)$ [38], [40], [41]. Assuming that rates $k_5$ and $k_6$ also apply to both Xe states [41], [42], and that the ratio $k_5/k_6$ is similar to $k_3/k_4$, we obtained values for $k_5$ and $k_6$ based on the recent scintillation measurements in six different Xe-$CH_4$ mixtures described in [43], [44]. We note that the resulting $k_5=2.5 \cdot 10^{-29}$ $cm^{-6}s^{-1}$ value is not very different from the $k_5=1.3 \cdot 10^{-28}$ $cm^{-6}s^{-1}$ value given earlier in [41] for $Xe(1s_4)$.

Finally, we note that although $CF_4$ itself may decay radiatively upon dissociative excitation of the molecules [46], [47], this does not contribute to EL in our simulations, because the process rarely occurs for the concentration and electric field ranges in the present work.

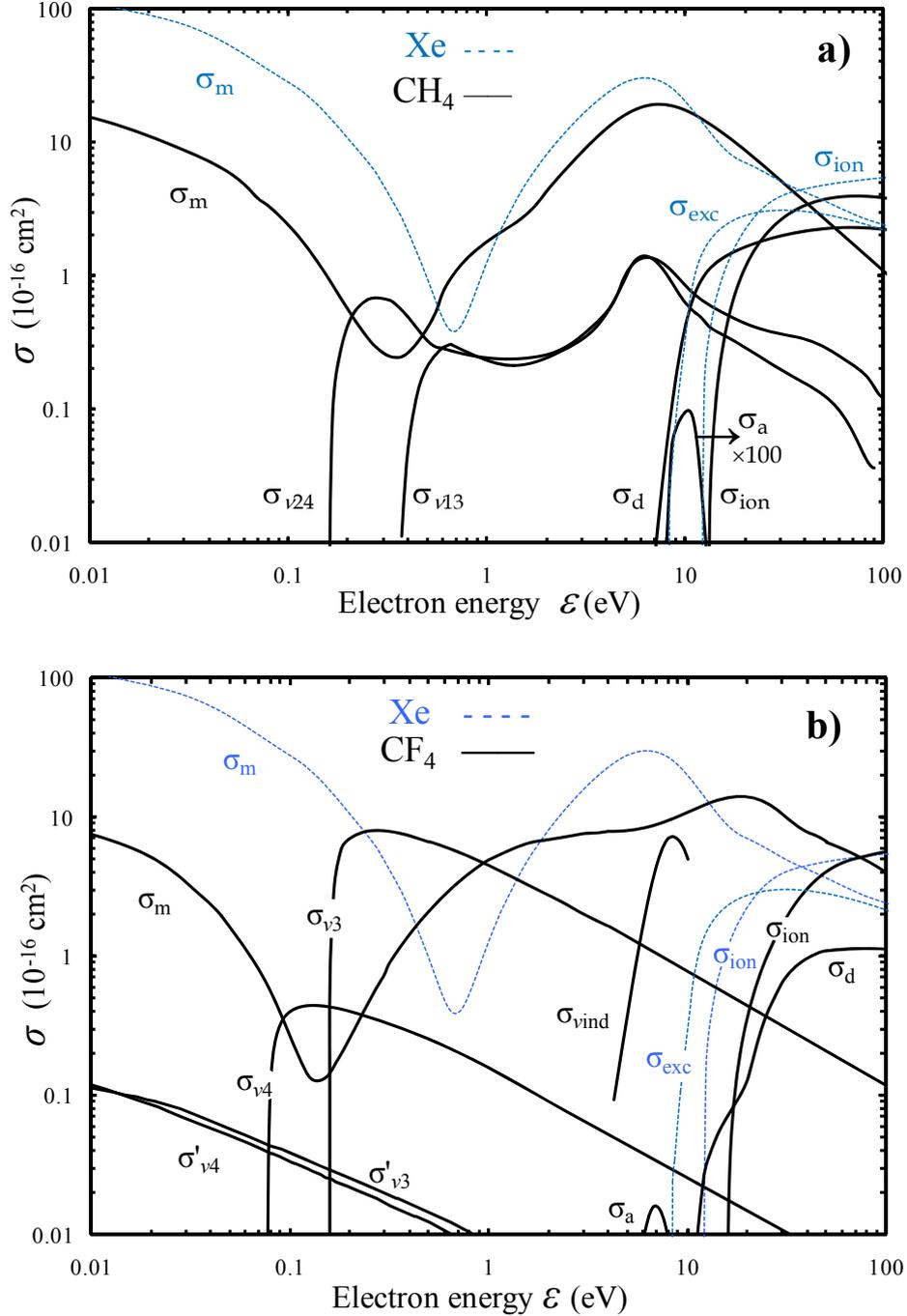

**Figure 1.** Electron scattering cross sections in a) Xe-$CH_4$ and b) Xe-$CF_4$ used in the Monte Carlo simulations: elastic momentum transfer ($\sigma_m$), vibrational excitation ($\sigma_v$), superelastic collisions ($\sigma'_v$ at $T$=293 K), electron attachment ($\sigma_a$), neutral dissociation ($\sigma_d$), electronic excitation ($\sigma_{exc}$), and ionization ($\sigma_{ion}$).

In pure Xe with planar geometry, absolute values for the EL yield $Y$ per unit length (number of scintillation photons produced per electron per cm) were obtained by Monte Carlo simulation in [15], where the results were described by the linear increase in Eq. (2) ($p$ in Torr and $E/p$ in V cm$^{-1}$ Torr$^{-1}$) or Eq. (3) ($N$ in cm$^{-3}$ and $E/N$ in Td)[2]

$$Y/p \text{ (cm}^{-1}\text{ Torr}^{-1}) = -a + b\, E/p = -a + b\, G \tag{2}$$

$$Y/N \text{ (}10^{-17}\text{ cm}^2) = -a' + b\, E/N = -a' + b\, S \tag{3}$$

with $a = 0.1325$, $b = 0.1389$ and $a' = 0.4020$, and where we denote the reduced electric fields as $G=E/p$ and $S=E/N$.

The above equations reproduce well the EL onset and the slope of the approximately linear increase with the electric field observed in [3] and in various later experiments (see [14], [20]), giving the value $G_0=E_0/p=0.95$ V cm$^{-1}$ Torr$^{-1}$ or $S_0=E_0/N=2.89$ Td for the EL threshold in Xe. For fields above this threshold but below electron multiplication, the number $\mathcal{H}$ of photons produced in planar geometry by one electron along an EL gap of length $D$ is then given by

$$\mathcal{H} = (-a + b\, G)\, p\, D \tag{4}$$
$$= (-a' + b\, S)\, N\, D$$

in terms of the reduced drift distance $pD$ (Torr cm), or $ND$ ($10^{17}$ cm$^{-2}$). Recent absolute measurements of EL in Xe have been made which are described by linear expressions very similar to the above [21].

As mentioned before, in the present work we have also investigated the EL output and fluctuations in pure Xe for the case of cylindrical geometry. The Monte Carlo model for the electron drift in this geometry has been described earlier in [35]. The main difference from planar geometry is in the equations that describe the motion of the electrons along the free paths between collisions under the radial increasing field, which are integrated in six-dimensional phase space by the "leap-frog" numerical method.

The Monte Carlo results for the number $\mathcal{H}$ of scintillation photons produced by an electron drifting in pure Xe towards the axial anode will be seen in Section 3.3 to approach closely the values calculated from the analytic expressions described in Eq. 5 or Eq. 6 below, which are derived from the integral $\mathcal{H} = \int Y(r)\, dr$ when we take for the local yield $d\mathcal{H} = Y(r)\, dr$ the expression from planar geometry ($\mathcal{H} = \int [-a + b\, G(r)]\, p\, dr$ for instance). The integration limits are $r=r_0$ and $r=r_a$, where the field at $r_0$ is the EL threshold in planar geometry ($G_0$, $S_0$), and the field at $r_a$ (anode surface) is a value ($G_a$, $S_a$) chosen just below the electron multiplication (Xe ionization) threshold. In terms of the reduced anode radius, $pr_a$ or $Nr_a$, or in terms of the reduced anode voltage $K = G_a\, p\, r_a = S_a\, N\, r_a$, the integration gives the linearly increasing behavior

$$\mathcal{H} = [a\,(1-G_a/G_0) + b\, G_a \ln(G_a/G_0)]\, p\, r_a \tag{5}$$
$$= [a'\,(1-S_a/S_0) + b\, S_a \ln(S_a/S_0)]\, N\, r_a$$

$$\mathcal{H} = [a\,(1/G_a - 1/G_0) + b \ln(G_a/G_0)]\, K \tag{6}$$
$$= [a'\,(1/S_a - 1/S_0) + b \ln(S_a/S_0)]\, K$$

---

[2] at $T=293$ K we have $p/N \cong 3.034\ 10^{-17}$, so $N$- and $p$-reduced variables are related by $ND$ ($10^{17}$ cm$^{-2}$) = $pD$ (Torr cm)/3.034, $Y/N$ ($10^{-17}$ cm$^2$) = 3.034 $Y/p$ (cm$^{-1}$ Torr$^{-1}$) and $E/N$(Td)=3.034 $E/p$ (V cm$^{-1}$ Torr$^{-1}$).

In the next Section, the Monte Carlo simulation results obtained in the present work will be described.

## 3. Results and discussion

### 3.1 Electron drift velocities $v_d$ and characteristic energies $\varepsilon_{kL}$ and $\varepsilon_{kT}$ in Xe, Xe-CH$_4$ and Xe-CF$_4$

In this Section, we present Monte Carlo electron drift parameters calculated in Xe-CH$_4$ and Xe-CF$_4$ mixtures with molecular concentrations relevant for the present work. In these calculations, a sample of $2\;10^4$ electrons with zero initial energy is followed in the gas at $p$=760 Torr and $T$=293 K for a drift time long enough to guarantee that equilibrium with the field is reached. We note that the drift parameters are pressure-independent as far as density effects related with multibody scattering can be neglected (below $p$~10 atm [48]-[49]). More extensive results in Xe-CH$_4$ and Xe-CF$_4$ mixtures at higher molecular concentrations and in the pure CH$_4$ and CF$_4$ molecular gases, can be found in [32]-[34], together with comparison with the experimental measurements available in the literature.

Fig. 2 and Fig. 3 represent the calculated electron drift velocities $v_d$ and characteristic energies $\varepsilon_{kL} = eD_L/\mu$ and $\varepsilon_{kT} = eD_T/\mu$ in Xe, Xe-CH$_4$ and Xe-CF$_4$ as a function of the reduced electric field $E/N$, where $D_L$ and $D_T$ are the longitudinal and transversal diffusion coefficients and $\mu = v_d/E$ is the electron mobility. Curves are shown for pure Xe and for Xe doped with CH$_4$ concentrations $\eta_{CH4}$=0.1%, 0.5% and 1% in Fig. 2, and for pure Xe and Xe doped with CF$_4$ concentrations $\eta_{CF4}$=0.01%, 0.05% and 0.1% in Fig. 3.

Fig. 2 and Fig. 3 show how the addition of CH$_4$ or CF$_4$ to Xe, even in very small concentrations, will significantly change the characteristics of the electron longitudinal and transversal diffusion in the gas, allowing a decrease in $\varepsilon_{kT}$ and $\varepsilon_{kL}$. On the other hand, the addition of CH$_4$ or CF$_4$ to Xe may increase the drift velocity $v_d$, which tends to go through a maximum and exhibit the negative differential conductivity effect (decrease of $v_d$ at increasing $E/N$) [29]-[34]. These effects are related to the efficient energy cooling of the electrons achieved at low electron impact energies by vibrational excitation collisions with the molecules (large cross-sections for vibrational excitation emerge in the region of the Ramsaeur-Townsend minimum of the elastic cross-sections, see Fig. 1).

In a gas detector where amplification is based on the production of EL in a scintillation region, the primary electrons produced by the ionizing event will first drift in the detector absorption region under an electric field just high enough to guide them towards the scintillation region but low enough to prevent electronic excitation altogether. In a Xe detector, this means that drift fields are chosen lower than ~2.5 Td (the inelastic threshold is ~3 Td as mentioned before). Looking at Fig. 2 and Fig. 3, we verify that this is the field range where doping Xe with CH$_4$ or CF$_4$ will enable drift parameters to be optimized by a judicious choice of mixture composition and electric field. However, this will always be achieved at the expense of lower EL yields and higher fluctuations as shown in next Section, and some compromise must be found.

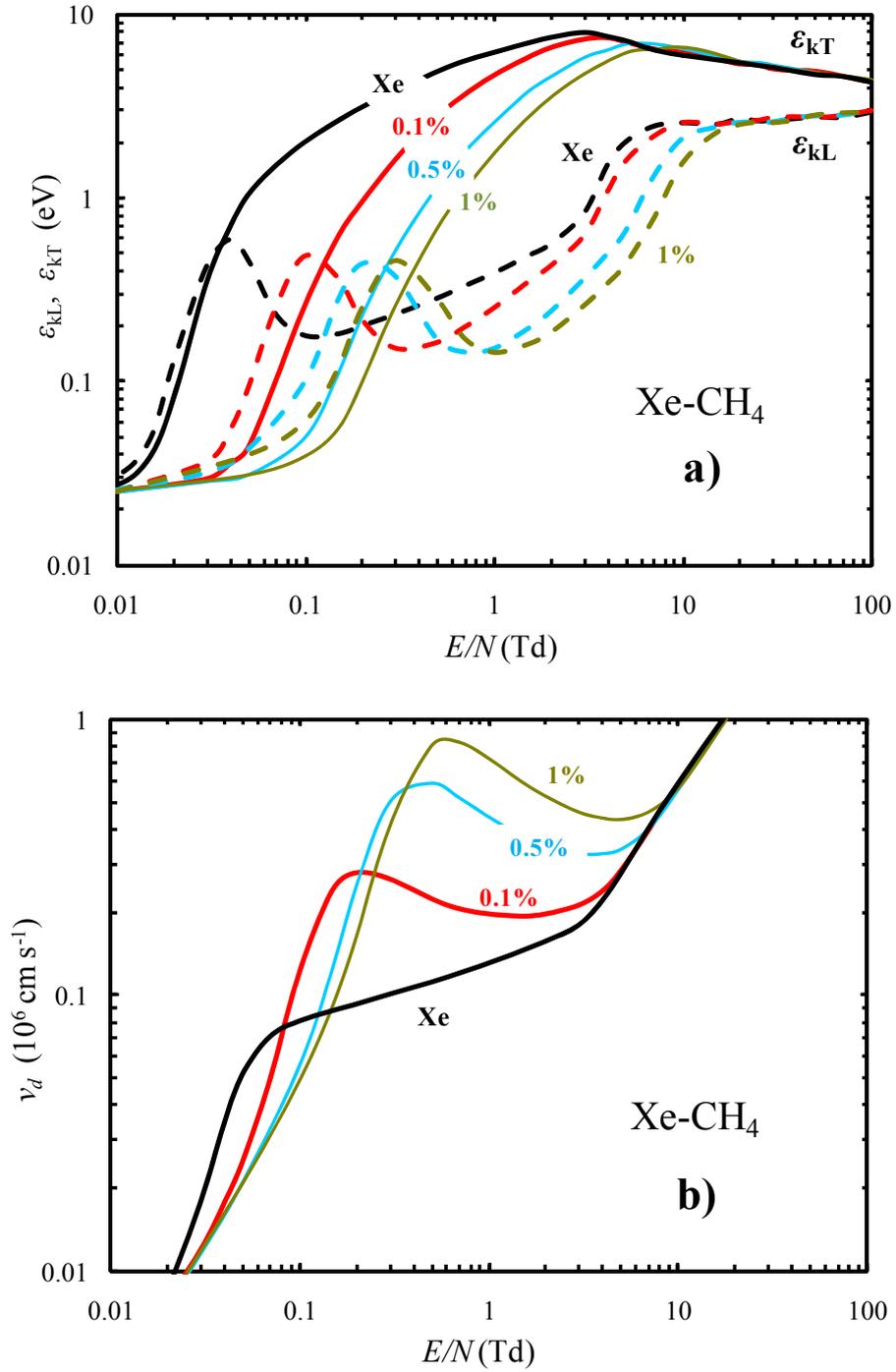

**Figure 2.** Monte Carlo results for a) characteristic electron energies $\varepsilon_{kL}$ and $\varepsilon_{kT}$ and b) electron drift velocity $v_d$ as a function of the reduced electric field $E/N$ in pure Xe and in Xe-CH$_4$ mixtures with CH$_4$ concentrations $\eta_{CH4}$=0.1%, 0.5% and 1%.

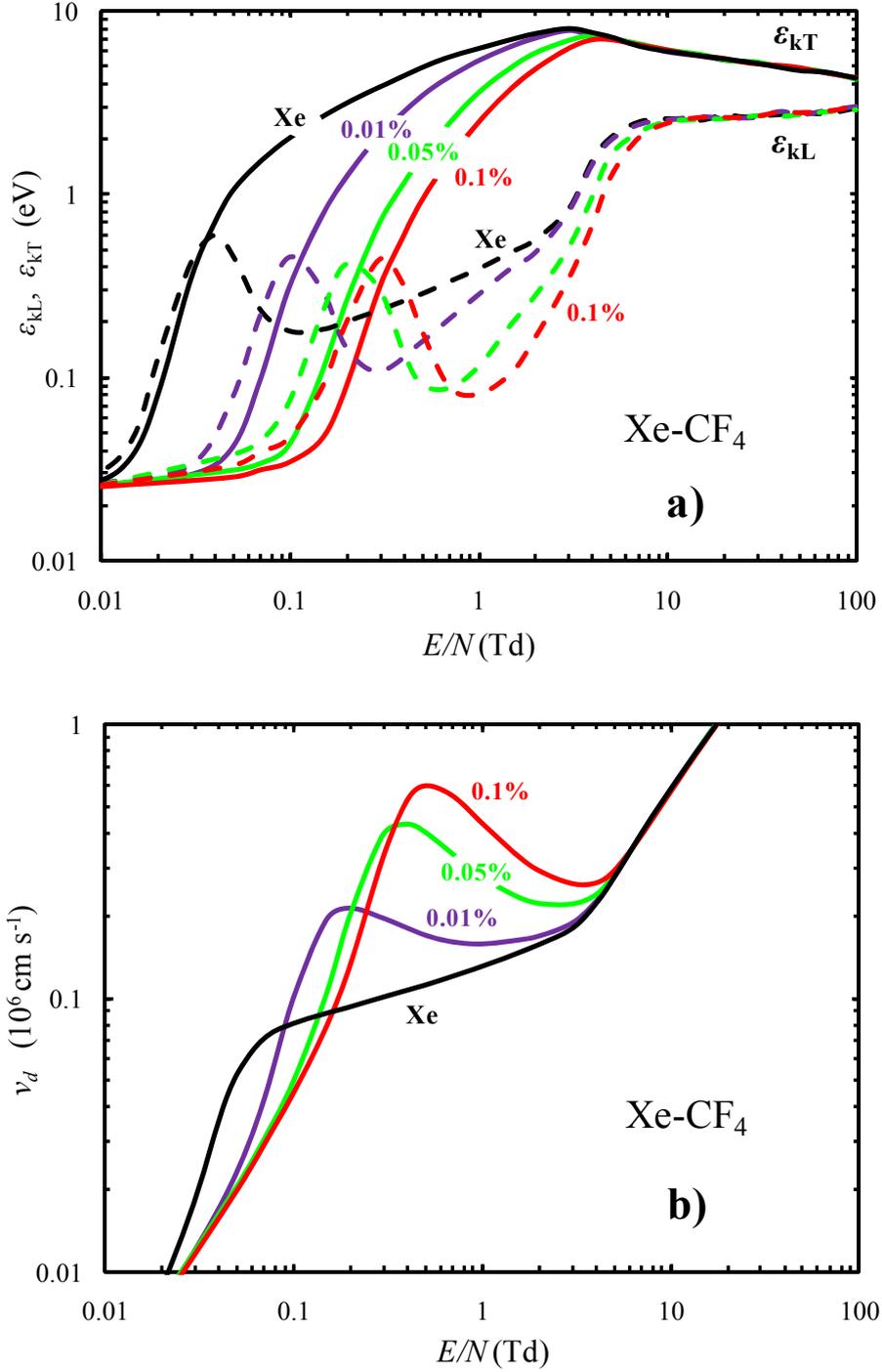

**Figure 3.** Monte Carlo results for a) characteristic electron energies $\varepsilon_{kL}$ and $\varepsilon_{kT}$ and b) electron drift velocity $v_d$ as a function of the reduced electric field $E/N$ in pure Xe and in Xe-CF$_4$ mixtures with CF$_4$ concentrations $\eta_{CF4}$=0.01%, 0.05% and 0.1%.

## 3.2 Electroluminescence and fluctuations in Xe, Xe-CH$_4$ and Xe-CF$_4$ in planar geometry

In this Section, we examine EL yields and their fluctuations in planar geometry in pure Xe and Xe doped with CH$_4$ or CF$_4$, following the drift of a sample of 5 10$^3$ electrons with thermal initial energy across a $D$=0.5 cm EL gap in the gas at $p$=7600 Torr (10 atm) and $T$=293 K (reduced distance $ND$=1252.47 10$^{17}$cm$^{-2}$, $pD$=3800 Torr cm) under the influence of an applied uniform electric field. We recall that the NEXT experiment will operate at this kind of pressure. Moreover, our simulations will not be valid above $p$~10 atm, since the electron-scattering cross-sections do not account for multibody scattering effects which then become important as we mentioned earlier.

For pure Xe and for the same Xe-CH$_4$ and Xe-CF$_4$ mixtures examined in Section 3.1 (three concentrations $\eta_{CH4}$ lower than 1% and three concentrations $\eta_{CF4}$ lower than 0.1%), Fig. 4 represents the Monte Carlo results for the number $\mathcal{H}$ of EL photons produced per single electron travelling across the EL gap in the gas under electric fields below electron multiplication threshold. In Fig. 5 the corresponding values for the ratio $Q=J/\mathcal{H}$ are plotted, where $J = \sigma_\mathcal{H}^2/\mathcal{H}$ is the relative variance of the EL distributions. We recall that the parameter $Q$ adds directly to the Fano factor $F$ to give the intrinsic energy resolution $R_{int}$ (see Eq. 1), so it is desirable that $Q$ remains significantly lower than $F$.

We verify in Fig. 4 that the Xe EL yield $\mathcal{H}$ is very sensitive to the presence of the molecular additives. In fact, although the molecular concentrations $\eta$ are low, a significant decrease in $\mathcal{H}$ is observed, and higher fields would be required to reach higher $\mathcal{H}$ values. The effect is much more pronounced for CF$_4$ than for CH$_4$ (note that the $\eta_{CF4}$ concentrations are lower than $\eta_{CH4}$ by one order of magnitude).

The decrease in EL is explained as follows. First, as already mentioned, electrons are efficiently cooled down in the mixtures due to vibrational excitation collisions with the molecules. This way, mean electron energies are lower in the mixtures than in Xe for the same applied electric field, contributing to a reduction of the number of Xe excitations and of $\mathcal{H}$ in the mixtures. Moreover, electron attachment by the molecules can play an important role in reducing EL, as further discussed below. A third source for EL decrease in the mixtures are the quenching mechanisms (2- and 3-body collisions involving the molecules, see Section 2.) that inhibit the formation of the Xe$_2^*$ excimers, which are the scintillation emitting species. As an illustrative example, simulation results for $E/N$=15Td indicate that, as compared to pure Xe, electron cooling, electron attachment and the EL-quenching processes contribute, respectively, about 14%, 4% and 82% to the decrease of $\mathcal{H}$ in Xe-0.1%CH$_4$ and about 9%, 91% and 0% in Xe-0.1%CF$_4$. These numbers show that the dominant effect is quenching in Xe-0.1%CH$_4$ and attachment in Xe-0.1%CF$_4$.

For the same mixtures as were examined in Fig. 4, we observe in Fig. 5 that the fluctuations parameter $Q$, negligible in pure Xe, becomes larger in the mixtures and increases with the additive concentrations $\eta$. However, even though the range of concentrations $\eta_{CF4}$ is an order of magnitude lower than the $\eta_{CH4}$ range, the $Q$ values are much higher in Xe-CF$_4$ than in Xe-CH$_4$.

From the results for $Q$ in Fig. 5, we can conclude that the energy resolution will not in principle deteriorate appreciably for Xe-CH$_4$ mixtures with concentrations $\eta_{CH4}$ within the low range 0.1% to 1%, since the corresponding $Q$ curves are seen to fall well below the Fano factor in Xe ($F_{Xe}$~0.17). On the contrary, we observe that $Q$ in Xe-CF$_4$ mixtures is very sensitive to $\eta_{CF4}$, and is already as high as ~1 for concentrations $\eta_{CF4}$ as low as 0.1%. In conclusion, CF$_4$ may be ruled out as an additive to Xe for uniform fields in the EL range[3], because the intrinsic energy resolution in Xe-$CF_4$ becomes too large and very sensitive to $\eta_{CF4}$. We note that the Fano factor in pure molecular gases will be higher than in xenon (for $CH_4$ a value ~0.28 was measured in [50]) but for the low molecular concentrations added to xenon in this work, the Fano factor in the mixtures is not expected to be significantly higher than in pure xenon.

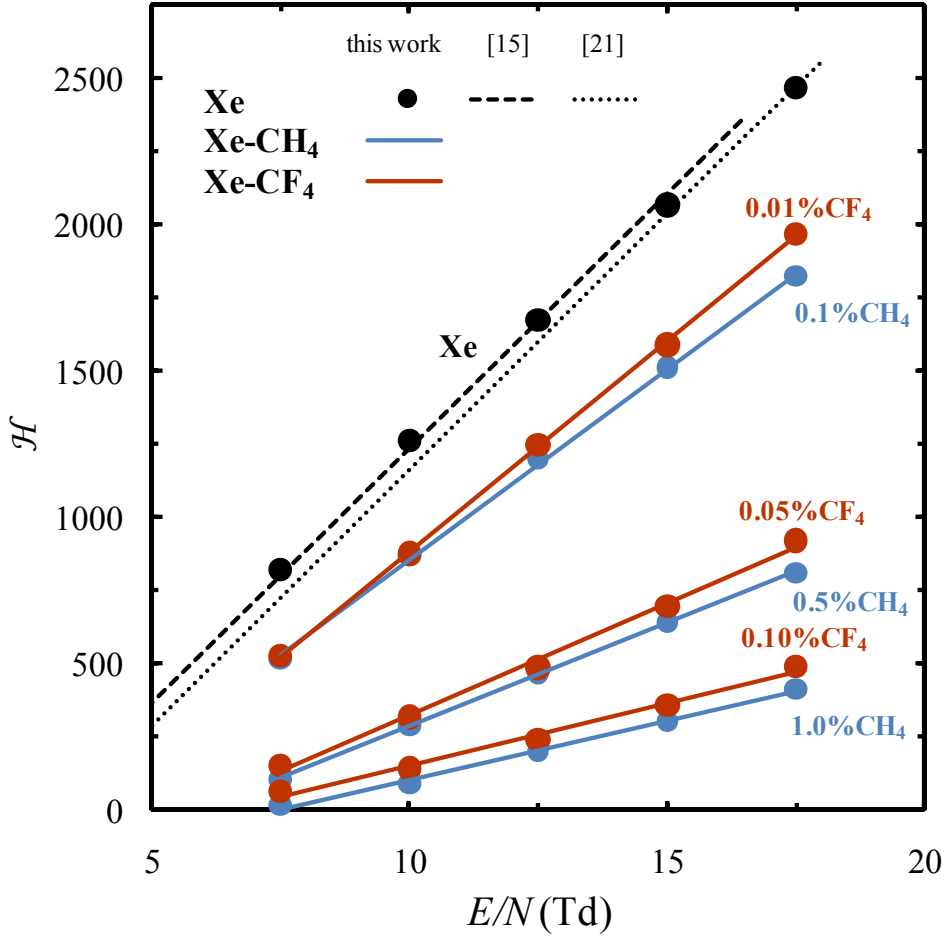

**Figure 4.** Monte Carlo results for the mean number $\mathcal{H}$ of EL photons produced in planar geometry under applied reduced electric fields $E/N$, when one electron drifts across a $D$=0.5 cm long scintillation region in Xe or in the Xe-$CH_4$ and Xe-$CF_4$ mixtures with the indicated $\eta_{CH4}$ and $\eta_{CF4}$ molecular concentrations at $p$ = 7600 Torr, T=293 K. The discontinuous lines represent the linear fittings found in [15] and in [21] to Monte Carlo and to experimental data, respectively.

---

[3] We note that the cross-section for electron attachment by $CF_4$ is narrowly peaked around 7 eV (Fig. 1b), so attachment is only important if electrons are allowed to drift long enough with energies close to this value. Electrons are known to survive attachment in standard cylindrical gas proportional counters (GPIC), which can operate with Ar-$CF_4$, Xe-$CF_4$ or even pure $CF_4$, as opposed to the high attachment fractions shown in Fig. 6 in Xe-$CF_4$ ($D$=5 mm, $p$=1 atm). This is because in GPICs the field rises rapidly over a few hundred microns close to the anode, so that electrons move quickly through the energy range of the attachment peak before reaching ionization energies, giving little chance for attachment.

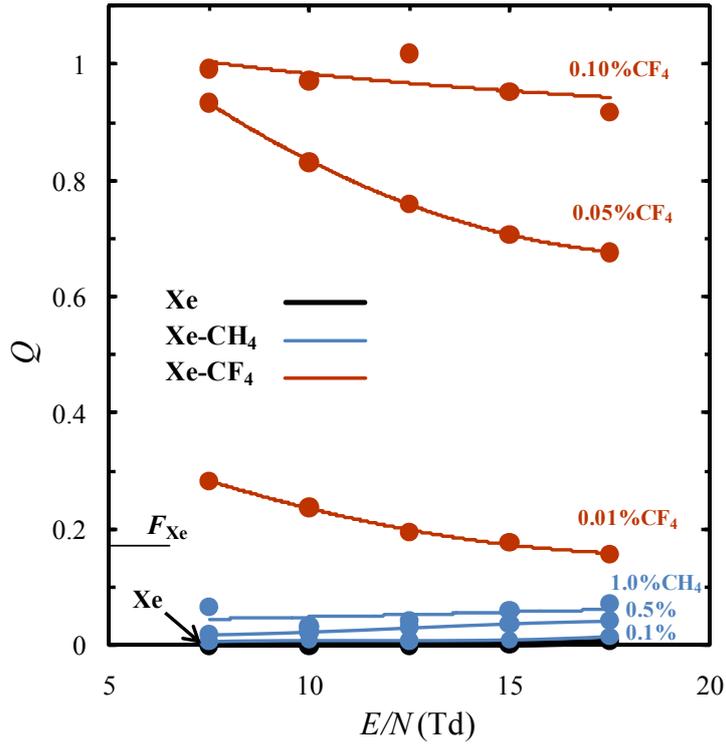

**Figure 5.** Monte Carlo results for the fluctuations parameter $Q=J/\mathcal{H}$, where $J$ is the relative variance $J=\sigma_\mathcal{H}^2/\mathcal{H}$, corresponding to the EL yield $\mathcal{H}$ results in Fig. 4. The bar labeled $F_{Xe}$ marks the Xe Fano factor.

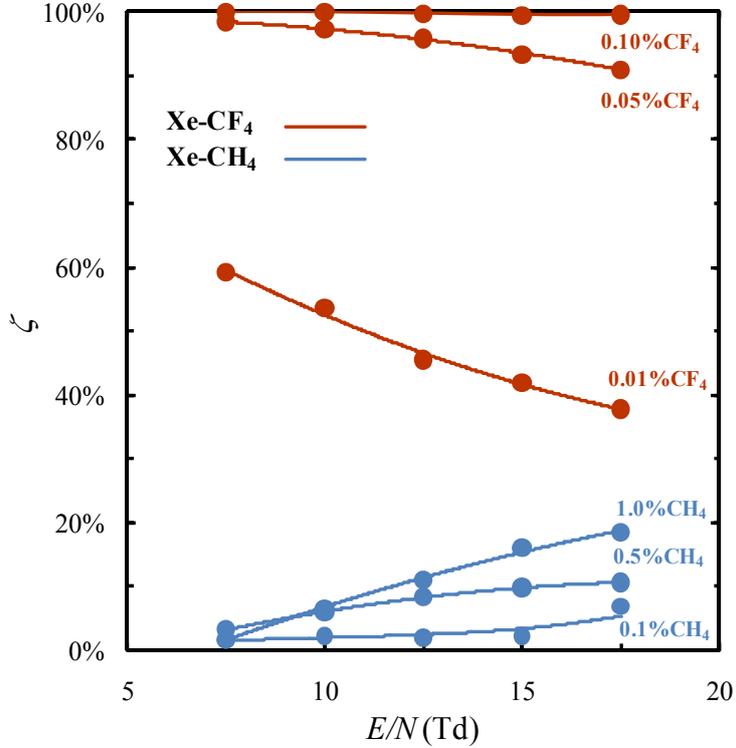

**Figure 6.** Monte Carlo results for the fraction $\zeta$ of electrons that become attached to $CH_4$ or $CF_4$ molecules when the EL yields $\mathcal{H}$ represented in Fig. 4 are obtained.

Finally, we plot in Fig. 6 the Monte Carlo results for the fraction $\zeta$ of electrons that become attached to the molecules somewhere along the EL gap, before they reach the anode at $D$. A comparison between Fig. 5 and Fig. 6 shows a clear correlation between the fluctuations parameter $Q$ and the $\zeta$ curves for attachment. The high $Q$ values in Xe-$CF_4$ reflect electron attachment, which obviously decreases the number $\mathcal{H}$ of EL photons produced in the gap and causes higher fluctuations $J$. The effect of attachment is much stronger for $CF_4$ than for $CH_4$, because the cross-section $\sigma_a$ for electron attachment by $CF_4$ is an order of magnitude higher than for attachment by $CH_4$ (see Fig. 1).

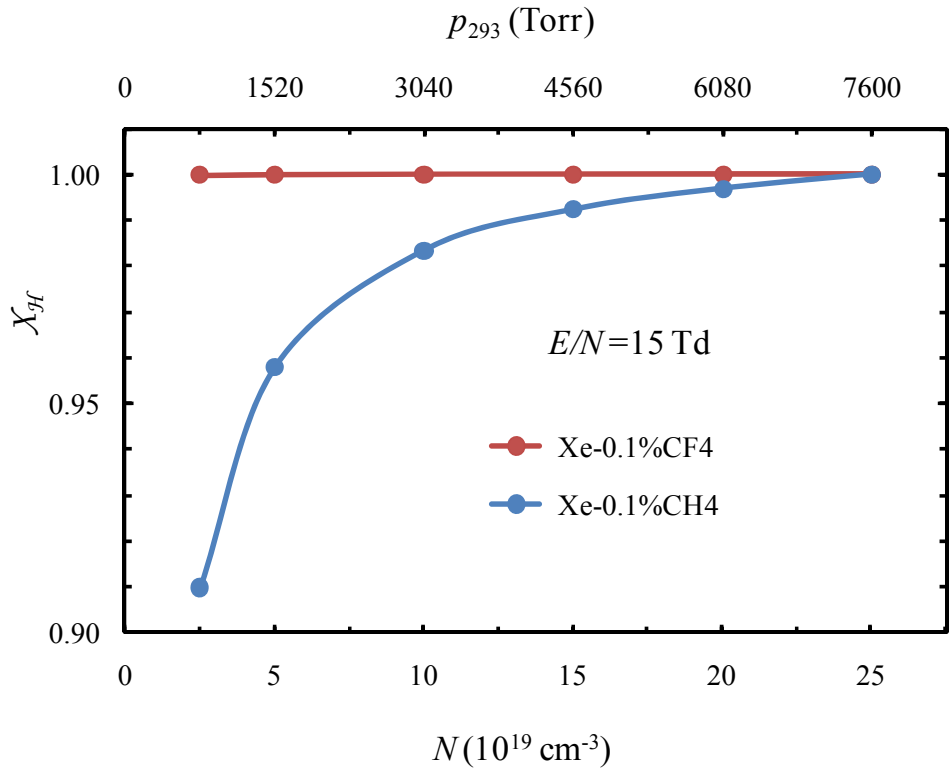

**Figure 7.** Monte Carlo simulation results for the EL relative yield $X_{\mathcal{H}}(p) = \mathcal{H}(p)/\mathcal{H}(7600)$ as a function of gas pressure in the range 760 to 7600 Torr (1 to 10 atm) at constant reduced distance $pD$ in the mixtures Xe-0.1%$CH_4$ and Xe-0.1%$CF_4$, for the applied uniform field $E/N$=15 Td.

The Monte Carlo results presented so far in this Section are calculated at $p$=7600 Torr ($p$=10 atm), but the EL yield $\mathcal{H}$ in the mixtures is expected to exhibit some pressure dependency, because the quenching mechanisms that reduce EL in the mixtures occur with probabilities that will vary with pressure. As an indication, normalized plots of $\mathcal{H}$ in Xe-0.1%$CH_4$ and Xe-0.1%$CF_4$ calculated at constant reduced distance $pD$ are shown on Fig. 7 within the pressure range 760 to 7600 Torr (1 to 10 atm), for a typical $E/N$=15 Td reduced applied field. In the Xe-0.1%$CF_4$ mixture, the curve is practically flat, because in this case EL reduction is mostly caused by electron attachment to $CF_4$, not by the pressure dependent quenching mechanisms. On the other hand, in Xe-0.1%$CH_4$ quenching dominates, and a clear pressure dependence is observed in Fig. 7. However, it was found that the parameter $Q=J/\mathcal{H}$ remains negligible and the pressure dependence of $Q$ is not significant. We note that, although different $pD$ values give distinct $\mathcal{H}$ curves, the normalization procedure used in Fig. 7 provided a description of pressure dependence by a unique curve for each mixture. Some dependence on $E/N$ might be expected, though, and the results shown correspond to $E/N$=15 Td. However, it was verified that the curves remain practically unchanged for $E/N$ values within the EL field range, which is short.

### 3.3 Electroluminescence and fluctuations in pure Xe: cylindrical geometry vs. planar geometry

In this Section, Monte Carlo simulation will be used to obtain and compare the fluctuations in the EL yield in pure Xe gas in planar and cylindrical geometries under charge-multiplication free regimes, investigating to what extent the fluctuations are increased in cylindrical geometry as compared to planar geometry, where they are known to be negligible and represent a major asset in standard Xe GPSCs as mentioned before. In the calculations, a sample of at least $10^3$ electrons with thermal initial energy are released in the gas at $p$=760 Torr, $T$=293 K from a radial distance $r$ large enough to guarantee that the field is lower than the Xe threshold for secondary scintillation (0.95 V cm$^{-1}$ Torr$^{-1}$=2.89 Td). The results in this section are summarized in Fig. 8, Fig. 9 and Fig. 10; they are for pure Xe, therefore they are not pressure dependent. Fig. 8 examines cylindrical geometry alone and compares Monte Carlo data for the EL yield $\mathcal{H}$ with results obtained from Eq. 5, while Fig. 9 and Fig. 10 compare Monte Carlo results for $\mathcal{H}$, $J$ and $Q$ in cylindrical and planar geometry.

Firstly, Fig. 8 shows the EL yields $\mathcal{H}$ in cylindrical geometry calculated from Eq. 5 for a field $S_a=(E/N)_a$=16 Td at the anode surface, together with curves for the reduced anode voltage $K=S_a N r_a$ and anode voltage $V_a= K \ln(r_c/r_a)$. For comparison, Fig. 8 includes five Monte Carlo data points for $\mathcal{H}$ (orange circles) calculated for the same $S_a$ and reduced anode radii values $Nr_a$=50.09, 75.14, 100.19, 125.24 and 150.28 $10^{17}$cm$^{-2}$ ($pr_a$=152, 228, 304, 380 and 456 Torr cm). The Monte Carlo calculations agree very well with the expected linear increase of $\mathcal{H}$ with reduced anode radius predicted by Eq. 5.

Secondly, Fig. 9 and Fig. 10 show Monte Carlo results for the EL yield $\mathcal{H}$ in cylindrical geometry (orange data), together with their relative variance $J = \sigma_{\mathcal{H}}^2/\mathcal{H}$ and the ratio $Q=J/\mathcal{H}$. In Fig. 9 $\mathcal{H}$, $J$ and $Q$ are plotted as a function of the reduced anode radius $Nr_a$ for the specific $S_a=(E/N)_a$=15 Td field at the anode surface, and in Fig. 10 they are plotted as a function of the field $S_a$ for a typical reduced radius $Nr_a$=75.14 $10^{17}$cm$^{-2}$ ($pr_a$=228 Torr cm), which corresponds for example to anode radii $r_a$=0.3, 0.06 or 0.03 cm if $p$=1, 5 or 10 atm, respectively.

For comparison, Fig. 9 and Fig. 10 include in addition Monte Carlo results for $\mathcal{H}$, $J$ and $Q$ in planar geometry (green data). To guarantee that $J$ and $Q$ are compared for similar EL yields $\mathcal{H}$ in both geometries, the reduced drift distances $ND$ in planar geometry at uniform fields $S_u=(E/N)_u=S_a$ are adjusted (as calculated from Eq. 4) to reproduce the EL yields $\mathcal{H}$ from each specified cylindrical geometry for the fields $S_a$ at the anode surface.

In Fig. 9, we observe that the relative variance $J = \sigma_{\mathcal{H}}^2/\mathcal{H}$ remains unchanged as $Nr_a$ or $ND$ increase. In fact, with no electron multiplication, $J$ is expected to be characteristic of the gas, EL electric field and geometry. Fig. 9 shows a $J$=0.088 mean value for the field $S_a$=15 Td in cylindrical geometry and the much lower value $J$=0.0052 for $S_u$=15 Td in planar geometry. On the other hand, the parameter $Q=J/\mathcal{H}$ is seen to decrease as $Nr_a$ or $ND$ increases, reflecting the increase in $\mathcal{H}$.

In Fig. 10, the Monte Carlo $\mathcal{H}$ results for the given cylindrical geometry are seen to increase with the field $S_a$ at the anode surface, as expected. This increase is faster than the characteristic linear increase with applied field in planar geometry observed in Fig. 4. In Fig. 10, in fact, the coincidence of the planar geometry $\mathcal{H}$ values with those obtained from cylindrical geometry could only be achieved by progressively increasing the drift distances in planar geometry as we moved to higher $S_u$ (more than threefold as we go from 6 to 17 Td), while the linear behavior in Fig. 4 applies for fixed reduced drift distance (Eq. 4).

In Fig. 10, we can compare the Monte Carlo results obtained for the fluctuation parameters $J$ and $Q$ in both geometries calculated at identical EL yields $\mathcal{H}$ at varying applied electric fields. As in Fig. 9 for the specific field 15 Td, the curves in Fig. 10 indicate that, for electric fields $S_a$ and $S_u$ within the EL range, $J$ and $Q$ increase by about one order of magnitude as we change from planar to cylindrical geometry. However, this is not expected to bring significant degradation to the intrinsic energy resolution $R_{\text{int}}$ (see Eq. 1), since we observe that $Q$ in the cylindrical geometry is still negligible compared to the Fano factor $F$ (about two orders of magnitude lower). We note that $Q$ will tend to be even lower at larger $Nr_a$ or $ND$ (see Fig. 9).

Thus, we may conclude that intrinsic fluctuations in $\mathcal{H}$ will give a negligible contribution to $R_{\text{int}}$ in both geometries, and identical $R_{\text{int}}$ values can in principle be reached. This is illustrated in Fig. 11, where $R_{\text{int}}$ is represented as a function of the x-ray energy $E_x$ absorbed in Xe gas. The two lower and practically coincident $R_{\text{int}}$ curves were calculated from Eq. 1 using the Monte Carlo $Q$ results from Fig. 10 at $S_a$=16 Td in cylindrical geometry and $S_u$=16 Td in planar geometry (when $\mathcal{H}$~150 EL photons), the Fano factor $F$=0.17 in Xe and the number of primary electrons per event $n=E_x/w$, where $w$=21.9 eV is the mean energy required to produce one primary electron in Xe [14] (neglecting recombination). However, the two $R_{int}$ curves are almost indistinguishable over the entire range of fields in Fig.10 (from 16 Td to 6 Td), because the parameter $Q$ remains well below the $F$=0.17 value over that range. We point out that $F$ and $w$ are asymptotic values reached for absorbed energies $E_x$ beyond the Xe photoionization absorption edges, since $F$ and $w$ for a lower $E_x$ range attain higher values and exhibit a discontinuous behavior with $E_x$, reflecting the absorption edges structure and Xe atomic shells as investigated earlier [5], [16], [17].

While the fluctuations in EL are found to give a negligible contribution to $R_{\text{int}}$ when electron multiplication is avoided, we find on the other hand that the accuracy of the geometry may be an important factor. This is shown in Fig. 11, where the curves $R'$ were calculated taking into account the additional fluctuations brought about by a 1% variation in the anode radius $r_a$ or in the drift distance $D$, at a constant applied anode voltage. We observe that, while at low incident energies the two sets of data are comparable, the curves $R'$ gradually deviate from $R_{\text{int}}$ at increasing $E_x$, showing that such uncertainty, albeit low, may lead to significant deterioration in the energy resolution for progressively higher energy events. The effect is seen to be more important in a cylindrical geometry, but on the other hand, highly accurate anode radii can be more easily achieved than parallelism between large area planar grids.

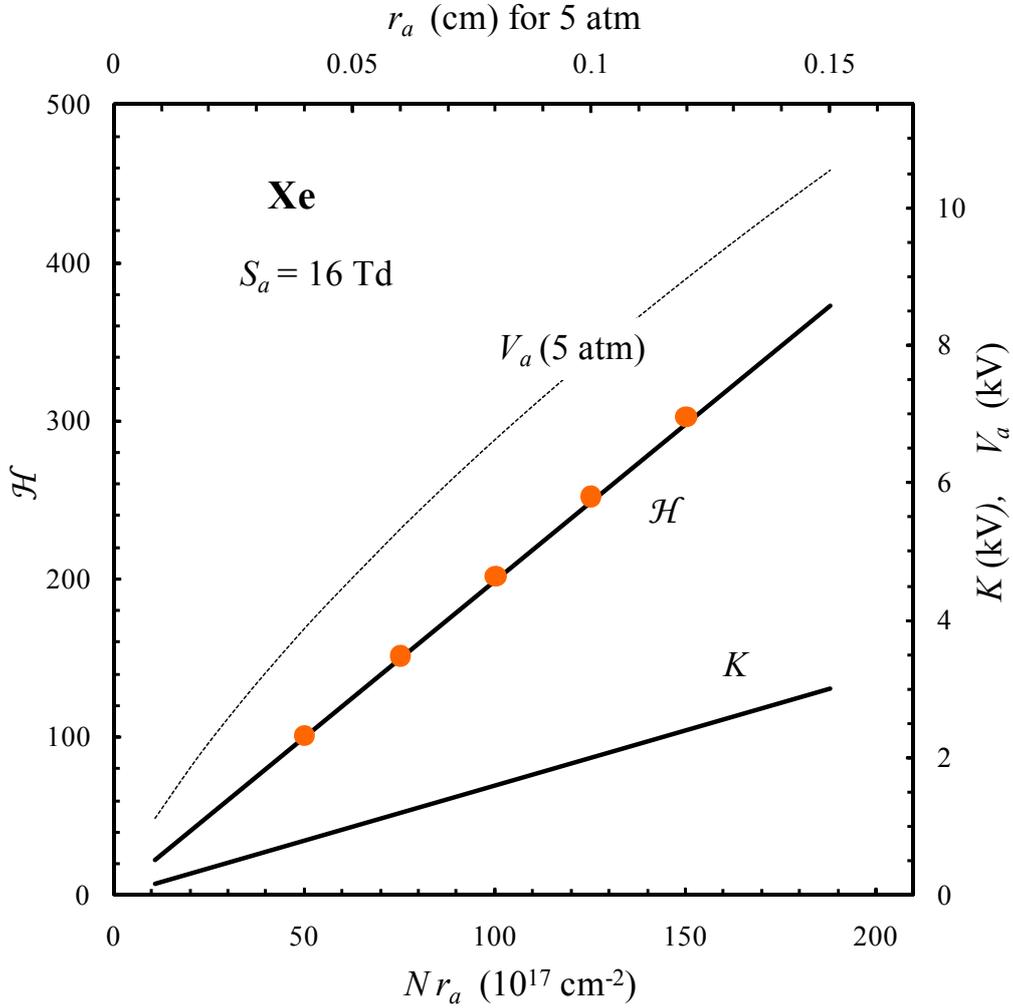

**Figure 8.** The continuous curves represent the EL yield $\mathcal{H}$ in Xe (mean number of EL photons produced per primary electron in the scintillation region) in cylindrical geometry as a function of the density-reduced anode radius $Nr_a$ in Xe calculated from Eq. 5 for a field $S_a=(E/N)_a=16$ Td at the anode surface, together with the corresponding reduced anode voltage $K=S_a N r_a$. The secondary horizontal axis indicates $r_a$ when $p=5$ atm, and the dashed curve is the anode voltage $V_a=K \ln(r_c/r_a)$ when, in particular, the chosen gas pressure and cathode radius are $p=5$ atm and $r_c=5$ cm. The data points marked as orange circles are calculated Monte Carlo results for the EL yield $\mathcal{H}$ at five distinct $Nr_a$ values, showing very good agreement with the linear increase of $\mathcal{H}$ with $Nr_a$ predicted by Eq. 5.

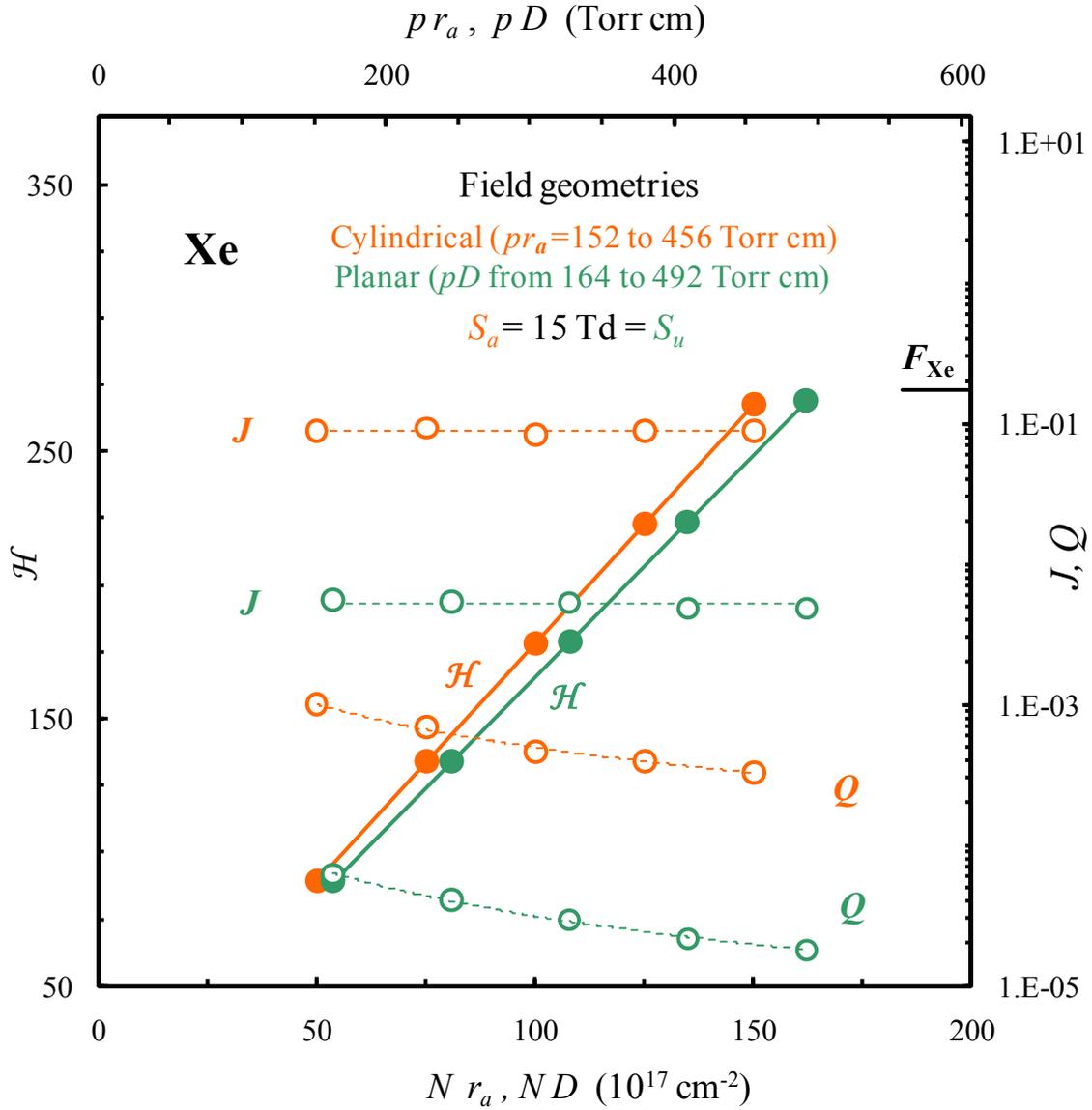

**Figure 9.** Monte Carlo simulation results for the electroluminescence yield $\mathcal{H}$ in Xe, the relative variance $J=\sigma_\mathcal{H}^2/\mathcal{H}$ and the ratio $Q=J/\mathcal{H}$ in cylindrical (orange) and planar (green) geometries. Orange data: $\mathcal{H}$, $J$ and $Q$ in cylindrical geometry as a function of the reduced anode radius $Nr_a$ for a field $S_a=15$ Td at the anode surface. Green data: $\mathcal{H}$, $J$ and $Q$ in planar geometry as a function of the reduced drift distance $ND$ for a $S_u=15$ Td uniform field. In planar geometry, the drift distances were adjusted to give the same yields $\mathcal{H}$ as in cylindrical geometry (points with same $\mathcal{H}$ in both geometries are distinct because in each case $ND$ is larger than $Nr_a$). The bar labelled $F_{Xe}$ marks the Fano factor in Xe.

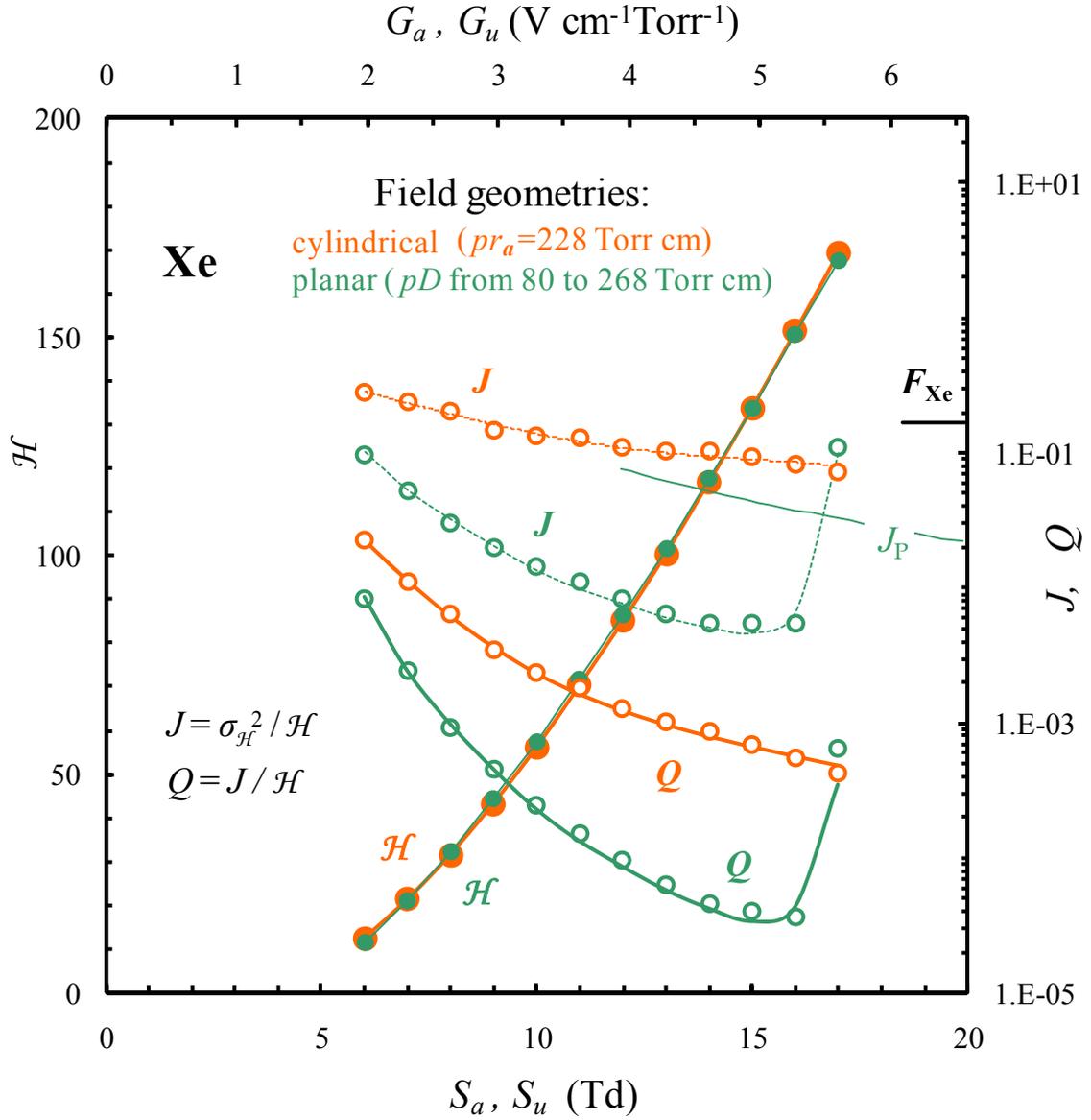

**Figure 10.** Monte Carlo simulation results obtained in Xe for the electroluminescence yield $\mathcal{H}$, the relative variance $J=\sigma_\mathcal{H}^2/\mathcal{H}$ and the ratio $Q=J/\mathcal{H}$ in cylindrical (orange) and planar (green) geometries. Orange data: $\mathcal{H}$, $J$ and $Q$ in cylindrical geometry as a function of the reduced field $S_a=(E/N)_a$ at the anode surface calculated for the reduced anode radius $Nr_a=75.14\ 10^{17}$ cm$^{-2}$ ($pr_a=228$ Torr cm). Green data: $\mathcal{H}$, $J$ and $Q$ in planar geometry as a function of the reduced uniform field $S_u=S_a$, for drift distances $pD$ which were adjusted to give the same yields $\mathcal{H}$ as in the specified cylindrical geometry. The jump in $J$ and $Q$ at the higher applied field in planar geometry is caused by electron multiplication, which did not occur in cylindrical geometry for the same field at the anode. The curve labeled $J_P$ is a rough approximation for $J$ in planar geometry found in [10]. The bar labeled $F_{Xe}$ marks the Fano factor in Xe.

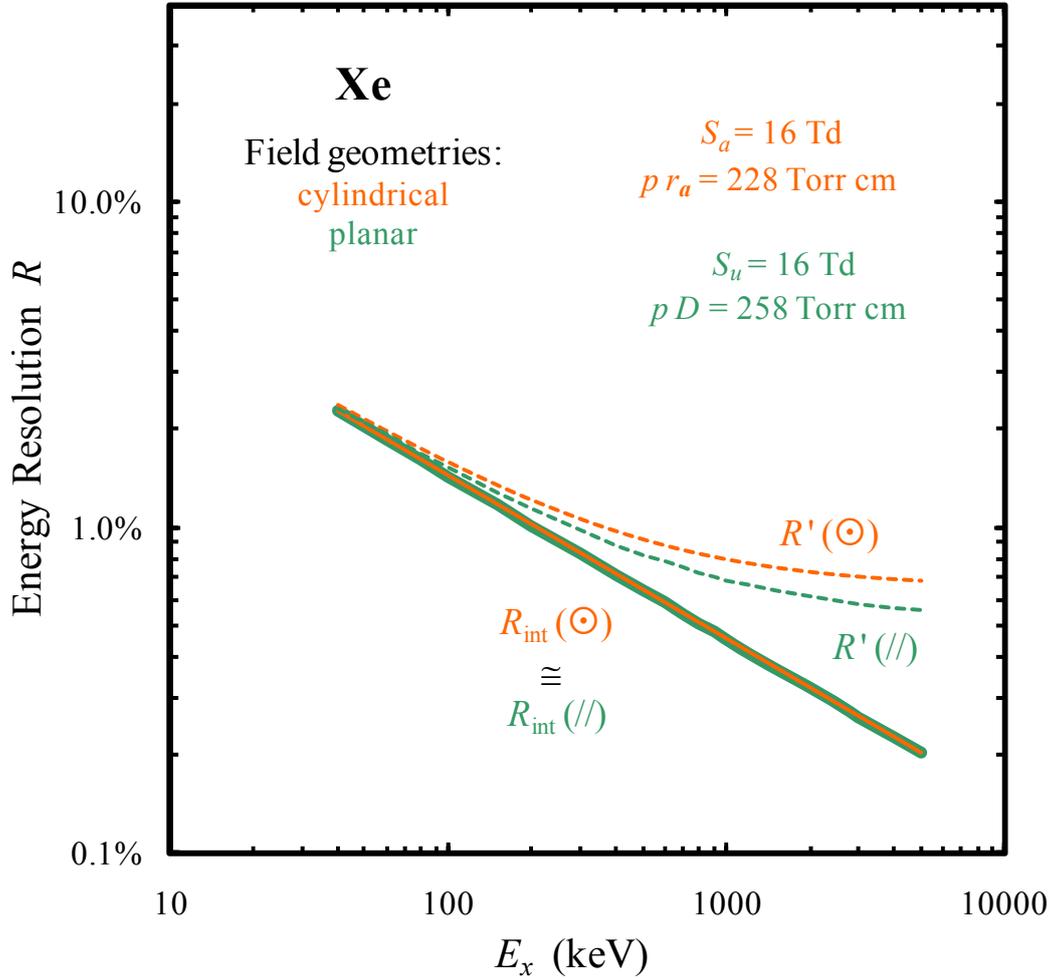

**Figure 11.** Continuous curves: intrinsic energy resolution $R_{int}$=2.35 sqrt[(1/$n$) ($F$+$Q$)] (Eq. 1) as a function of absorbed x-ray energy $E_x$ in Xe, where $n$=$E_x$/$w$, $F$=0.17, $w$=21.9 eV and $Q$ are the Monte Carlo values from Fig. 10 at $S_a$=16 Td in cylindrical geometry and $S_u$=16 Td in planar geometry (corresponding to $\mathcal{H}$~150 EL photons). The $R_{int}$ curves appear coincident for the two geometries because the $Q$ values are negligible compared to $F$. Dashed curves: 'intrinsic' energy resolution $R'$ when a term is added to $R_{int}$ taking into account the fluctuations resulting from a 1% uncertainty (standard deviation) on the anode radius in cylindrical geometry or on the drift distance in planar geometry at constant applied voltage. This term becomes predominant for $E_x$ above about 200 keV.

## 4. Conclusions

The addition of either $CH_4$ or $CF_4$ to Xe produces an increase in electron drift velocity and a reduction of electron diffusion, where both effects are related to a cooling of the electrons through vibrational excitation of the molecules. For Xe-filled gas detectors, doping with these additives may be advantageous when electrons must travel long drift distances through a large detector absorption region before reaching some kind of amplification stage. In some large detector applications, as planned for the NEXT experiment for the study of the double beta decay of $^{136}$Xe, amplification will be based on the production of electroluminescence (EL) in Xe in a regime below the charge multiplication threshold, in order to guarantee that EL fluctuations are minimal. However, the presence of molecular additives reduces the EL yield and increases its fluctuations, because a broader variety of collision processes become available for the electrons, such as vibrational excitation of the molecules and electron attachment, and also because alternative deactivation processes for the Xe excited states arise in the mixtures as they collide with the molecules, which result in quenching of the Xe scintillation emission. For the range of *E/N* investigated, these processes were found to play an important role in Xe-$CH_4$, while in Xe-$CF_4$ electron attachment is the dominant effect.

Using a standard uniform field planar geometry, a Monte Carlo simulation study was made of the EL yield and its fluctuations when Xe is doped with $CH_4$ or $CF_4$. The results indicate that doping Xe with $CH_4$ concentrations in a range up to ~1% may improve the drift parameters, as long as drift fields in the detector are chosen near *E/N* ~1 Td (Fig. 2). Depending on the intensity of the drift field, a higher or lower $CH_4$ concentration within that range will be appropriate. In addition, for the (higher) field range *E/N*~3 to 16 Td typical of the scintillation region of a Xe gas detector with amplification based on EL (a range between the Xe excitation and ionization thresholds), it was shown that the EL fluctuations parameter *Q* in these Xe-$CH_4$ mixtures can be kept well below the Xe Fano factor (*F* values for the mixtures with low additive concentrations as used in this work are not expected to be much higher than *F* in Xe), avoiding significant degradation of the intrinsic energy resolution $R_{int}$ (Fig. 5 and Eq. 1). This way, Xe-$CH_4$ mixtures with concentrations lower than 1% may be regarded in principle as having advantages over pure Xe as the gas filling for large gas detectors.

In contrast, the calculations show that doping Xe with $CF_4$ may not be a good choice. Although a reduction in electron diffusion may also be achieved in Xe-$CF_4$ at low fields (Fig. 3), we verify that, even for minute $CF_4$ concentrations, the EL yield is strongly reduced and the EL fluctuations term *Q* becomes very large and well above the Fano factor (Fig. 4 and Fig. 5). This happens essentially because, throughout the EL electric fields range, electron attachment to $CF_4$ molecules becomes much more important than to $CH_4$ molecules (Fig. 6), resulting in a much larger decrease and much higher fluctuations in the EL yield in Xe-$CF_4$ than in Xe-$CH_4$.

On the other hand, accurate parallelism between grids of the scintillation gap in planar geometry is difficult to achieve when detectors are very large. For that reason, large detectors may instead use multiwire cylindrical geometries for EL production. This is a possible choice for the case of the NEXT experiment, which has considered the production of EL around a series of parallel wires in scintillation regions at the two ends of the detector.

With this in mind, in the present work we have also investigated the fluctuations of EL produced in a cylindrical geometry in pure Xe, considering the drift of electrons towards a central wire at whose surface the electric field will not rise above multiplication threshold. Results are described in Fig. 9 and Fig. 10, showing that although the relevant fluctuation parameters are about an order of magnitude larger than in a planar geometry, they can still be negligibly small compared to the Fano factor and will not jeopardize energy resolution (see Fig. 11). Calculations show in addition that, to avoid degradation of the energy resolution, wire radius uniformity better than 1% should be achieved even for moderate high energy events (above ~200 keV).

In the future, our calculations may be extended to other Xe-based mixtures as well as Argon-based mixtures. Experimental measurements are under way to measure the EL yield and energy resolution of gas proportional-scintillation counters filled with the kind of gas mixtures examined in the present work.

## Acknowledgments

This work was carried out at Physics Department, University of Coimbra, Portugal, and was supported by the FEDER/QREN/POFC program through FCT (Fundação para a Ciência e Tecnologia, Portugal) project PTDC/FIS/112272/2009. J. Escada acknowledges support from FCT Grant SFRH/BD/22177/2005. We thank David Nygren for suggesting this work and for useful discussions.